# Framework and Classification of Indicator of Compromise for physics-based attacks

Vincent Tan
*Dissruptive Techologies Office*
*Home Team Science and Technology*
*Agency (HTX)*
Singapore
Vincent_TAN@htx.gov.sg

*Abstract*—Quantum communications are based on the law of physics for information security and the implications for this form of future information security enabled by quantum science has to be studied. Physics-based vulnerabilities may exist due to the inherent physics properties and behavior of quantum technologies such as Quantum Key Distribution (QKD), thus resulting in new threats that may emerge with attackers exploiting the physics-based vulnerabilities. There were many studies and experiments done to demonstrate the threat of physics-based attacks on quantum links. However, there is a lack of a framework that provides a common language to communicate about the threats and type of adversaries being dealt with for physics-based attacks. This paper is a review of physics-based attacks that were being investigated and attempt to initialize a framework based on the attack objectives and methodologies, referencing the concept from the well-established MITRE ATT&CK, therefore pioneering the classification of Indicator of Compromises (IoCs) for physics-based attacks. This paper will then pave the way for future work in the development of a forensic tool for the different classification of IoCs, with the methods of evidence collections and possible points of extractions for analysis being further investigated.

## I. INTRODUCTION

Today, industries and governments are constantly working on ways to safeguard sensitive information of the public and national interests. This led to the exploration of quantum communications which are based on the law of physics to protect the sensitive information. These physics laws of quantum mechanics allow particles, commonly known as qubits to transmit information along optical cables and any attempt to intercept the information will result in a collapse of the quantum state, thus leaving tell-tale signs of such activities. With such promises, there have been growing interests to integrate quantum-based technologies into existing communications infrastructure, such as QKD. The implications of such adoption were studied, physics-based vulnerabilities existed, and exploitations were demonstrated [1]. However, there is a lack of a framework that provides a common language to communicate about such threats and the type of adversaries being dealt with.

This paper is to first summarize the exploitation of the possible physics-based vulnerabilities that have already been investigated and attempts to initialize a framework based on the attack objectives and methodologies, referencing the concept from the well-established MITRE ATT&CK and finally, pioneering the classification of IoCs for physics-based attacks, which is the focus of this paper. An indicator in this paper is an artifact that can be observed, measured, or collected with the current state of technologies. Hence, IoC is a malicious indicator indicating that the quantum link or component might have been compromised.

The attempt to initialize a framework is referenced to the design and philosophy of the well-established MITRE ATT&CK [2], which is a well-known documented knowledge base of attacker tactics and techniques based on real-world observations created by MITRE in 2013. ATT&CK laid the foundation for the development of specific threat models, methodologies, and mitigations for the cybersecurity community globally. It provides a common language for both offense and defense, such that became a tool for different cybersecurity disciplines to communicate about threat intelligence, perform testing through red teaming, evaluate current defenses, track attacker groups and enhance network and system defenses against attacks.

The proposed initial framework in this paper is to demonstrate the possibility of referencing to well-established framework in cybersecurity, such as ATT&CK, and introduce new forms of considerations due to the emerging quantum communication technologies. This proposed initial framework will therefore simplify and bridge the quantum communications and cybersecurity communities towards a common goal in security.

The motivation and focus of this paper are to enable the future work of developing forensic tools for the different classification of IoCs, with the methods of evidence collections and possible points of extractions for analysis being further investigated.

This paper is organized as follows. In Section II, the review of physics-based attacks was done and categorized into attack objectives. With each attack objectives identified, different methodologies of the attacks were briefly discussed. In Section III, an attempt to map the ATT&CK model relationships with attacks based on quantum communication technologies was proposed. In Section IV, the classification of the IoCs for physics-based attacks was proposed to form a basis for the development of specific forensic tools in future for the different classification of IoCs. The paper is then concluded in Section V.

## II. ATTACK OBJECTIVES AND METHODOLOGIES

Many studies, experiments and demonstrations have been performed to show the exploitation of the physics-based attacks. In each attack, the attacker has to decide the entry points. These entry points can be categorized into attack objectives as goal to enter through (i) environment, (ii) source of photons, (iii) detectors of photons. Such attack objectives can be expanded in the future with more advanced attacks being discovered.

With the attack objectives identified, different methodologies can be classified accordingly to detail what the attack is and/or how the attack can take place. These different methodologies are briefly discussed in each of the attack objectives.

*A. Environment*

From the environment, we identified two possible sources of attacks that could occur in a quantum network, which is either implemented either using fibre optic links or free-space links. Firstly, tapping of quantum links on fibre optics when the attackers have access or knowledge to the location of the fibre optics. This is similar to the classical tapping of fibre optics and commercial providers are already demonstrating on the ease of such tapping of fibre optics, assuming having knowledge of the location of fibre optics. Hence, this will not be further discussed in this paper.

Secondly, through free-space quantum links. There are two possible modes of physics-based attacks that have been studied/simulated for free-space links. It is possible to conduct an "in-Field-Of-View (FOV)" attack and prior studies [3] showed that it is more relevant to conduct an "Out-of- FOV" attack on the Optical Ground Stations (OGSs) using a laser to induce a temporary or permanent quantum denial of service (DoS). The focus of such attack is to have intense illumination that caused temporary or permanent harm to optical or electronic components. An example of such attack is optical jamming within "in-FOV" where such attack was experimented and mitigation technique was proposed. [4]

*B. Source of Photons*

Quantum attacks at the source of photons have been heavily studied and some being demonstrated on commercial QKD systems. There are many examples of possible attacks at source, such as photon-number splitting, phase-remapping, trojan horse, Faraday mirror, laser seeding. [1] One of the well-known quantum attacks at source is photon number splitting (PNS) attack and targets at imperfect photon source. Due to the weak coherent pulses generated by laser used in QKD implementations, there will be non-zero probability of multiple-photon pulses. This is because the photon number of a phase-randomized weak coherent pulse follows the Poisson distribution. Therefore, an attacker can exploit the multi-photon pulses and launch the PNS attack.

Phase re-mapping attack is another well-known quantum attack at source and one of those demonstrated on commercial QKD systems with Eve using the same set-up as Bob to launch attack. Time delay between reference pulse and signal pulse was shifted by adding variable optical delay line and re-mapping the phase small enough into the low quantum bit error rate (QBER) range. This experiment showed that only a QBER of 19.7% was introduced, which is below the well-known 25% error rate for an intercept-and-resend attack in BB84.

*C. Detectors of Photons*

Similarly, attacks at the detectors of photons were investigated extensively [1], and detectors are generally deemed as more vulnerable as compared to source of photons. The commonly known theoretical attacks are double-click, fake-state, memory attacks for example. Some of the experimented attacks are time-shift and detector-control attacks. Dectector-control attacks can be generally classified into three types (i) detector-blinding attack, where attacker sends bright light to detectors; (ii) detector-after-gate attack, where attacker sends multi-photon pulses at the position after the detector gate; (iii) detector-super-linear attack, where attacker exploits the superlinear response of single photon detectors during the rising edge of the gate. Time-shift attacks has also been successfully demonstrated on commercial QKD systems. Such attacks do not make any measurements on the quantum state, and thus the quantum information is not destroyed.

## III. MAPPING OF ATT&CK MODEL RELATIOSHIPS

The design and philosophy of the well-established MITRE ATT&CK. [2] is based on a well-documented knowledge base of cyber attacker tactics and techniques based on real-world observations. The ATT&CK model has the core components of (i) Tactics, (ii) Techniques, (iii) Sub-techniques and (iv) Documented adversary usage of techniques, their procedures, and other metadata.

In a form of simplified explanation, (i) Tactics, refer to the "why" of the technique which is the attack's tactical objective. (ii) Techniques, refer to "How" the attacker has achieved and/or "What" the attacker has done to achieve the tactical objective. (iii) Sub-techniques, refer to the more detailed and specific means by which the tactical objectives are achieved. (iv) Documented adversary usage of techniques, their procedures, and other metadata refer to the steps on the specific implementation that attacker has used for techniques and sub-techniques.

As extracted from the ATT&CK technical paper [2], an ATT&CK object model relationship is presented in a visual form as shown on Fig. 1.

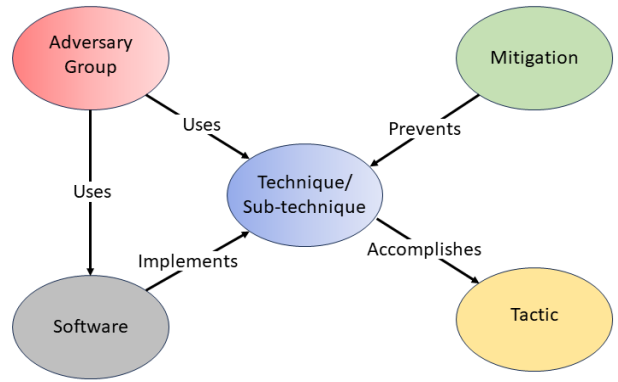

Fig. 1. ATT&CK Model Relationships

Referencing to ATT&CK framework, this paper will attempt to map the attack objectives and methodologies as discussed in Section II. The generic mapping is presented in Fig. 2. Hence, with the generic mapping in Fig. 2, the feasibility of mapping is further demonstrated in Fig. 3, Fig. 4 and Fig. 5, using different attack objectives and methodologies. For each of the examples demonstrated, the tools used to implement the attack methodology and mitigation to prevent the attack are only briefly discussed as these are not the focus of this paper. For adversary group, this will be appropriate in future when documentation of such attacks is available. Hence, will not be discussed in this paper.

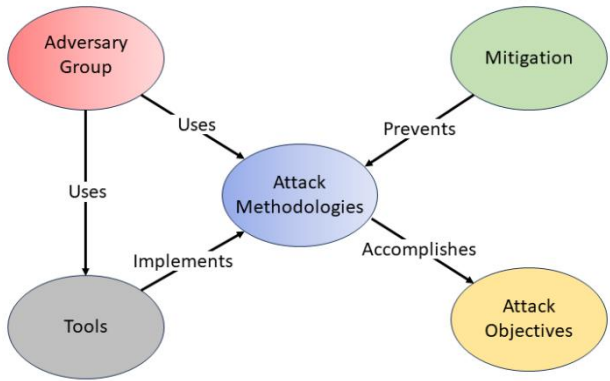

Fig. 2. Mapping the ATT&CK model relationships with attacks based on quantum communication technologies using attack objectives and methodologies.

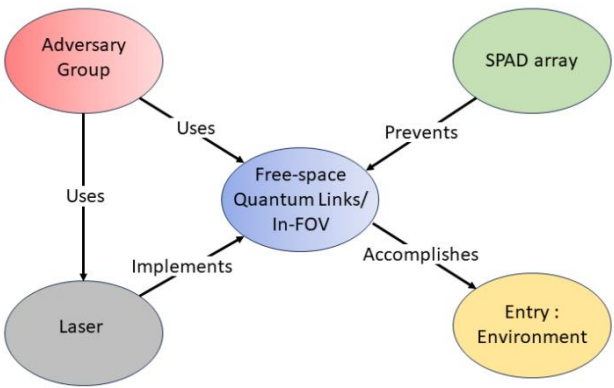

Fig. 3. Mapping the ATT&CK model relationships with attack objective as goal to enty by environment.

In Fig. 3. the attack objective is to enter by the environment and accomplishes this goal by making use of the free-space quantum links to perform optical jamming within "in-FOV". The attack makes use of laser as a tool to implement the attack and prior studies [4] experimented that it is possible to prevent such attacks with single photon avalanche detector (SPAD) array.

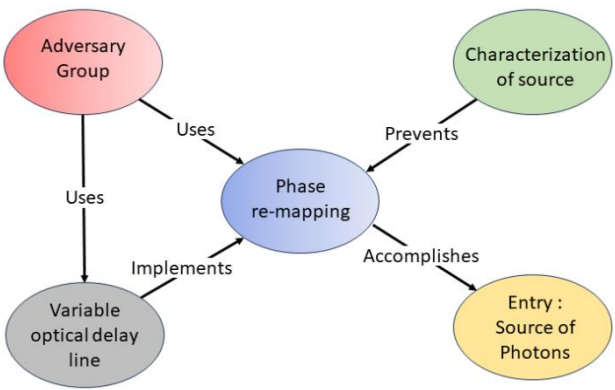

Fig. 4. Mapping the ATT&CK model relationships with attack objective as goal to enty by source of photons.

In Fig. 4. the attack objective is to enter by the source of photons and accomplishes this goal by re-mapping the phase small enough into the low QBER range. The attack makes use of the variable optical delay line to implement the shift of time delay between the reference pulse and the signal pulse, and then re-mapping the phase as experimented in prior studies [5]. It has been proposed to prevent such attacks with characterization of source [6].

In Fig. 5. the attack objective is to enter by the detector of photons and accomplishes this goal by first blinding the detectors, then sends a bright pulse with tailored optical power. Consequently, an intercept-and-resend attack without increasing QBERs can be launched. The attack makes use of bright light to implement the attack as demonstrated in various prior studies [7], [8]. It has been proposed to prevent such attacks by improving the optical scheme of the decoding unit in the QKD, such that using a coupler with an asymmetric splitting ratio [9] to distinguish the detection characteristic of the SPD with blinding attack against without blinding attack.

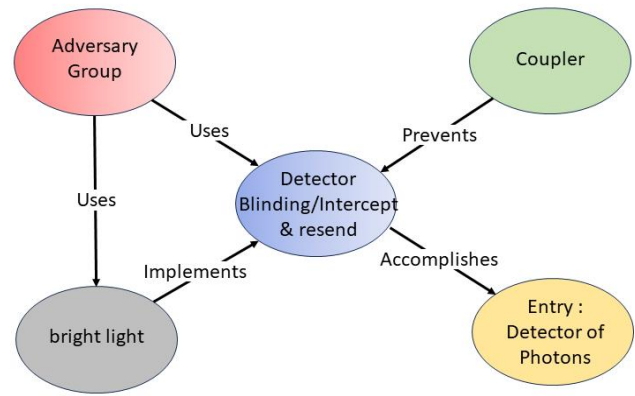

Fig. 5. Mapping the ATT&CK model relationships with attack objective as goal to enty by detectors of photons.

It has thus been demonstrated on the feasibility of mapping ATT&CK model relationships with attack objectives and methodologies that are based on quantum communication technologies.

IV. CLASSIFICATION OF INDICATOR OF COMPROMISES

The focus of this paper is to classify the IoCs for physics-based attacks that can be observed /collected /measured due to the attacks at environment, source and detectors of photons, which are the attack objectives discussed in Section II. The usefulness in mapping of ATT&CK model discussed in Section III will be seen in this work on classification of IoCs of physics-based attacks. During an investigation, with the mapping presented in Fig. 2, an investigator will be able to reference and understand the tools used by attacker to implement the attack methodologies to accomplish the attack objectives, therefore the type of IoC to be observed/ collected/ measured will be more targeted and streamlined. The investigator will thus focus on the IoCs that will likely result from the tools used and specific attack methodologies used, as shown in Fig. 6.

With the work on classification of IoCs, this will form a basis for the development of specific forensic tools for the different classification of IoCs in future. This work should be further expanded when more attacks are discovered and new methods of observing, collecting and measurement of the compromises are discovered. For now, this work will focus on three classifications of IoCs: (i) Quantum Bit Error Rate (QBER), (ii) Real-time characterization (iii) Received power.

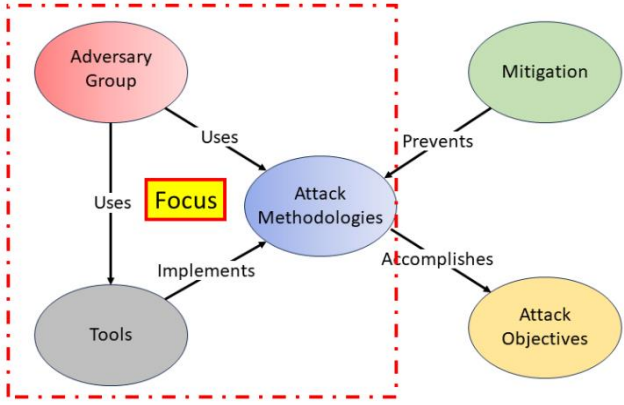

Fig. 6. Tools and specific attack methodologies used by attacker will likely result in IoCs and therefore, the mapping helps investigation work to be more targeted and focus. This demonstrated the benefits of the proposed framework that is referenced to ATT&CK.

*A. QBER*

Most of the studies used measurable parameters like key rate or QBER to analyze if the quantum communication is secured or possibility of being attacked. QBER is defined as:

$$\text{QBER} = \frac{\text{number of error bits}}{\text{total number of bits transmitted}} \quad (1)$$

QBER is a measure of the errors that occurs in quantum communication and typically expressed in percentage. Depending on the application, the acceptable range of QBER values may vary. For example, for a QKD system based on BB84 protocol, an intercept-and-resend attack will induce a typical well-known 25% error rate. Prior studies [3] indicated that a practical QBER threshold should commonly be around 8 – 12%. Anything beyond this threshold range, it indicates an abnormality and thus, a possible attack. Hence, it is worthwhile to note that while QBER should be kept as low as possible or according to the baseline set per application, when there is a sudden spike in QBER, this observation should be investigated as it may be an indication of an attack attempt.

*B. Real-time characterization*

- Afterpulse probability: It has been demonstrated [10] that real-time monitoring of single photon detectors for different ratios of intercepted per transmitted photons is effective to detect attacks such as after-gate attack. This was done by comparing to the standard operating mode and a measurable increase in the afterpulse probability that reveals an attack.

- Deadtime: By measuring the time intervals between consecutive counts generated by a non-gated Si detector, other than able to characterize a continuous running detector, the deadtime can be monitored in real-time as well [10]. The deadtime which are multiples of the analog to digital converter (A/D) sampling period when compared to manufacturer's specification, if there is a notable gap in the deadtime, this reveals an attack such as blinding attacks.

- Photocurrent: In prior studies [11], a photocurrent monitor circuit is used to output a photocurrent monitor readout value as a function of a constant APD photocurrent through a few steps of processing and conversion. The photocurrent monitor readout value is expected not to exceed 8100 for single sampled values and any value larger indicates an attack.

- Photon statistics: Measurement and characterization of photon statistics at source enables detection of attacks. [6] This is possible because weak coherent source follows Poisson distribution and the characterization of photon statistics at source enables estimation of any information leakage due to multi-photon pulses. An inaccurate measurement of the mean photon number may lead to unaccounted information leakage, thus an undetected attack.

*C. Received power*

Measurement of the received power ($P_{recv}$) at the single photon detector (SPD) is a good indicator of compromise for attacks using lasers as investigated in prior studies. [3] It has been established that for $P_{recv} \geq 10^{-15}$ W, the noise level will be too high and thus, indicative of an attack. This received power threshold is established based on the limits of the number of noise or background photons where the count of noise or background is at $3 \times 10^{-5}$ per pulse to maintain an error rate of below 5%.

When $P_{recv} \geq 10^{-3}$ W, it will likely cause thermal blinding to the semiconductor-based avalanche photodiodes (APDs). The damages will get more severe as the received power increases. Up to the point of $P_{recv} \geq 10^3$ W, general melting is likely to start.

It is worthwhile to note that for QKD systems, there will be a few optical components such as bandpass filters or polarizing beam splitters placed before the SPDs, thus effectively reducing the received laser power to the SPDs.

V. DISCUSSION AND FUTURE WORK

There have been many prior studies done on physics-based attacks on quantum communications. Some of these physics-based attacks are theoretical and many have been demonstrated. With the experiments of different physics-based attacks, mitigation countermeasures were also gradually proposed and studied. However, there is a lack of common language to communicate about such threats and the type of adversaries being dealt with. This caused a gap between the quantum communications and cybersecurity communities.

This paper therefore attempts to bridge the gap between the quantum communications and cybersecurity communities by referencing to well-established MITRE ATT&CK, which is a well-known documented knowledge base of attacker tactics and techniques based on real-world observation and initialize a framework.

The framework first categorized the attack objectives by entry points; (i) environment, (ii) source of photons and (iii) detectors of photons. The attack objectives and methodologies to achieve these objectives were discussed briefly and mapped to the ATT&CK model that has the core components of (i) Tactics, (ii) Techniques, (iii) Sub-techniques and (iv) Documented adversary usage of techniques, their procedures, and other metadata.

Presented in Fig. 2, the ATT&CK model relationships were successfully mapped generically with attacks based on quantum communication technologies using attack objectives and methodologies. It was then further demonstrated on the feasibility of the mapping to different attack objectives and methodologies using an example from each of the different attack objectives such as entry through environment, source of photons and detectors of photons as shown in Fig. 3, Fig. 4 and Fig. 5 respectively.

With the proposed initial framework, the IoCs of the physics-based attack methodologies were classified. The classification of these IoCs is an important step to frame the critical design parameters of an operational forensic tool which may be specific for the different classification of IoCs. This forensic tool to be developed in future is to empower an investigator with the ability to extract and collect evidence for analysis of the physics-based attacks and ready for the quantum era in the near future.